\documentclass[aps,showpacs,amsmath,amssymb]{revtex4}

\usepackage{graphicx}
\usepackage{dcolumn}
\usepackage{bm}

\newcommand{\beq}{\begin{equation}}
\newcommand{\eeq}{\end{equation}}
\newcommand{\beqa}{\begin{eqnarray}}
\newcommand{\eeqa}{\end{eqnarray}}
\newcommand{\beqar}{\begin{eqnarray*}}
\newcommand{\eeqar}{\end{eqnarray*}}
\newcommand{\bra}[1]{\mbox{$\left\langle{#1}\right|$}}
\newcommand{\ket}[1]{\mbox{$\left|{#1}\right\rangle$}}

\def\I{{\rm i}}

\newcounter{saveeqn}

\begin{document}

\title{Direct diagonalization of Fock Space \\ for an exact solution of pairing model}
\author{An Min Wang}\email{anmwang@ustc.edu.cn}
\affiliation{Quantum Theory Group, Department of Modern Physics\\
University of Science and Technology of China, Hefei 230026,
People's Republic of China}

\begin{abstract}
We investigate the exact solution of BCS pairing model using direct
diagonalization of Fock space. By the data analysis and numerical
calculation, we verify the symmetry between energy spectrum of Fock
subspaces, obtain the common structure features of energy gaps and
energy bands in Hamiltonian spectrum of reduced model, propose the
formula to estimate the lowest energy levels in all of the subspaces
of reduced model, and suggest a scheme to estimate the respective
energy spectrum which can reveal the structure of energy spectrum of
pairing model.
\end{abstract}


\maketitle

Recently, the Bardeen, Cooper and Schrieffer (BCS) model for
superconductivity \cite{bcs} has been connected with the problems in
different areas of physics such as superconductivity, nuclear
physics, physics of ultrasmall metallic grains and color
superconductivity in quantum chromodynamics \cite{report,dukelsky}.
Since Ricardson's works from 60's \cite{richardson} to now , the
exact solution of the reduced (constant) BCS pairing model has been
well known. However, for the more general cases in which the
coupling coefficients among different pairs are different, the exact
solution of BCS pairing model is on studying
\cite{Dukelsky2,Amico1,Amico2,Volya}. In fact, the exactly solvable
models have been proven to be very useful tools to understanding the
physics of strongly correlated many-body quantum systems. There are
a serial of the important and interesting works about the pairing
model being published that have not cited here (see the references
in \cite{report,dukelsky}).

Let us consider the BCS pairing Hamiltonian \cite{report}:%
\begin{equation}
H_{\mathrm{BCS}}=\sum_{m=1}^{L}\frac{(\varepsilon_{m}-\varepsilon_{F})}%
{2}(n_{m}+n_{-m})-\sum_{m,l=1}^{L}V_{ml}c_{m}^{\dag}c_{-m}^{\dagger}c_{-l}c_{l}
\label{bcsh}%
\end{equation}
where $n_{\pm m}\equiv c_{\pm m}^{\dagger}c_{\pm m}$ are the
electron number operators, $c_{m}^{\dagger}(c_{m})$ is the
fermionic creation (annihilation) operator, and the coupling
coefficients are real and symmetric, that is $V_{ml}=V_{lm}$.
Based on the Refs.\cite{report,dukelsky,simulation}, one is able
to study equivalently its spin-analogy form as the following
\begin{equation}
H_p^{(L)}=\frac{1}{2}\sum_{m=1}^{L}\epsilon_{m}\sigma_{z}^{(m)}%
-\frac{1}{2}\sum_{m<l=1}^{L}V_{ml}\left(\sigma_{x}^{(m)}\sigma_{x}^{(l)}%
+\sigma_{y}^{(m)}\sigma_{y}^{(l)}\right)  \label{hspin}%
\end{equation}
where $\epsilon_{m}=(\varepsilon_{m}-\varepsilon_{F})-V_{mm}$, and
the constant term $\sum_{m}\epsilon_{m}/2$ has been ignored, which
vanishes anyway since we cut off symmetrically above and below
$\varepsilon_{F}$. Note that $L$ is the number of the pairs or
qubits here.

Quantum theory tell us, it is natural to exactly solve the pairing
model by the direct diagonalization in Fock space. Several works
have been published in this aspect, for example
\cite{Burglin,Molique}. However, its feasibility still has to be
further considered under no any approximation. Clear calculations
and interesting analysis of the results need to further study.
Obviously, a difficulty of exact solution of spin-analogy of pairing
model is, in the large-$L$ limit, the obsession of exponentially
complicated problem to the direct diagonalization. Fortunately, this
difficulty can be partially avoided by making use of decomposition
of Fock space. At least, for some given subspaces of Fock space, the
direct diagonalization becomes a polynomial problem. On the other
hand, the direct diagonalization needs that one knows the general
forms of the Hamiltonian in the given subspaces. This implies that a
key problem is to seek for their explicit forms. As soon as they are
found and expressed, it is indeed feasible to exactly solve the
pairing model by the direct diagonalization of Fock space.

Actually, our motivation is arose originally by our study on quantum
simulation of pairing model on a quantum computer
\cite{simulation,Ourqst,Ourqse}. Because we did not know how to
implement the relevant approximations in the simulating solution of
pairing model in a quantum computer, and we realize that the direct
diagonalization will be feasible and direct at least on a quantum
computer, and so we would like to develop it for the quantum
simulation of pairing model in the near future.

Now we first recall the technology of the decomposition of Fock
space and give out a strict and simple proof about it. In fact, the
spin space $S_{\rm spin}^{(L)}$ in a $L$-pair (qubit) system can be
divided into the different subspaces which correspond to the
different numbers of spin-up states, that is $S_{\rm
spin}^{(L)}=S_{0}^{(L)}\oplus S_{1}^{(L)}\oplus
S_{2}^{(L)}\oplus\cdots\oplus S_{L}^{(L)}$, where the subspace $n$,
$i.e$ $S_n^{(L)}$, is a subspace with $n$ spin-up states $\ket{0}$
(corresponding to ``occupation"). Thus, we can use
$i_1,i_2,\cdots,i_n$ to indicate the values of positions appearing
$0$ in the bit-string $\alpha_1\alpha_2\cdots\alpha_L\;(\alpha_i=0$
or $1$), and then the bases of $S_n^{(L)}$ can be denoted by
$\ket{s^{(L)}_{i_1i_2\cdots i_n}}\; (n\neq0)$ and
$\ket{s_0^{(L)}}\in S_0^{(L)}$. Obviously, the dimension of
$S_n^{(L)}$ is $\displaystyle\frac{L!}{n!(L-n)!}$.

From
$\left(\sigma_x^{(m)}\sigma_x^{(l)}+\sigma_y^{(m)}\sigma_y^{(l)}\right)
=\left(\sigma_x^{(m)}+\I\sigma_y^{(m)}\right)
\left(\sigma_x^{(l)}-\I\sigma_y^{(l)}\right)=\sigma_{m}^{(+)}\sigma_{l}^{(-)},
(m\neq l)$ in eq.(\ref{hspin}), it follows that for the arbitrary
basis $\ket{s_{i_1\cdots i_n}^{(L)}}$ belonging to $S_n^{(L)}$,
$H_p\ket{s_{i_1\cdots i_n}^{(L)}}$ also belongs to $S_n^{(L)}$
because that $\sigma^+$ and $\sigma^-$ appear in pairs or do not
appear in the various terms of $H_p$. It implies that
$\bra{s_{i_1\cdots i_{m}}^{(L)}}H_p\ket{s_{i_1^\prime\cdots
i_n^\prime}^{(L)}}=0$, (If $m\neq n; m,n=1,2,\cdots, L$) and
$\bra{s_0^{L}}H_p\ket{s_{i_1\cdots i_n}^{(L)}}=0$. Therefore, we
have proved that the pairing model Hamiltonian is able to be
decomposed into the direct sum of submatrices in Fock subspaces,
that is
\begin{equation}
H_{p}^{(L)}=H_{{\rm sub}0}^{(L)}\oplus H_{\rm sub1}^{(L)}\oplus
H_{\rm sub2}^{(L)}\oplus\cdots\oplus
H_{{\rm sub}L}^{(L)} \label{dirsum1}%
\end{equation}
It is clear that two the simplest eigenvectors of $H_p^{(L)}$ are in
the one dimension $S_0^{(L)}$ and $S_L^{(L)}$ respectively with the
eigenvalues $\mp\sum_{m=1}^L\epsilon_m/2$. Thus, the concerned
subspaces or submatrices of Hamiltonian in our method only includes
those from $1$ to $L-1$.

Furthermore, we can derive out \beqa \label{hpaction}
H_p^{(L)}\ket{s^{(L)}_{i_1i_2\cdots i_n}}
&=&\left(-\frac{1}{2}\sum_{m=1}^L\epsilon_m+\sum_{a=1}^n
\epsilon_{i_a}\right)\ket{s^{(L)}_{i_1i_2\cdots i_n}}
-\sum_{b=1}^n\sum_{m<i_b, m\neq i_a,
a<b}^{i_b-1}V_{mi_b}\ket{s^{(L)}_{P^+_{i_1 i_2\cdots
i_n;i_b;m}}}\nonumber\\
& & -\sum_{a=1}^n\sum_{l>i_a, l\neq i_b, a<b}^{L}V_{i_a
l}\ket{s^{(L)}_{P^+_{i_1 i_2\cdots i_n;i_a;l}}}, \quad (i_a, i_b\in
\{i_1,i_2,\cdots,i_n\})\eeqa where $P^+_{i_1 i_2\cdots i_n;i_c;r}$
is such a permutation that $i_c$ is dropped, $r$ is added, and
$\{i_1,i_2,\cdots,i_{c-1},i_{c+1}$, $\cdots,i_L,r\}$ is rearranged
according to their values from small to large. In terms of the
orthogonality of bases, we obtain immediately the general and
explicit forms of Hamiltonian submatrices of pairing model in the
Fock space. Consequently, we can carry out the direct
diagonalization of them within the limitation of the computer's
power.

\begin{figure}[h]
\begin{center}
\includegraphics[scale=0.60]{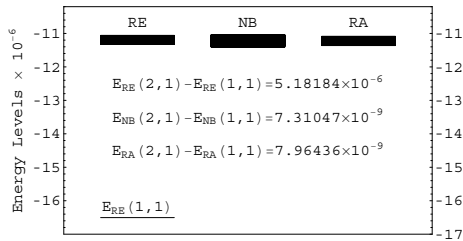}
\end{center} \vskip -0.2in
\caption{The energy levels in the three kinds of models RE, NB and
RA ($L=170$, $\lambda=20$)}\label{mypic1} \vskip -0.1in
\end{figure}

It must point out that in the half filling case $i. e.$
$n=\displaystyle\left[L/2\right]$, the dimension of its submatrix
$H_{\rm subhf}^{(L)}$ is the highest and arrives at
${L!}/{\left([L/2]![(L+1)/2]!\right)}$, where ``$[M]$" means taking
the integer part of $M$. Note the fact that $H_{\rm subhf}^{(L)}$ is
a sparse matrix, for example, the dimension of $H_{\rm sub8}^{(16)}$
is $12870$, but the number of its nonzero elements is only $65\times
65$, one is able to diagonalize it to some $L$. Nevertheless, with
$L$ increasing, the dimension of half filling submatrix will exceeds
quickly the limits of large scale diagonalization in a classical
computer. Perhaps, quantum simulation can solve this difficulty
\cite{simulation,Ourqst,Ourqse}. Here, we focus on the problem how
obtain the more knowledge of Hamiltonian spectrum for the moderate
$L$, which is helpful for understanding many body quantum theory, as
well as for providing a precision comparison with the possible
result of quantum simulating in the near future. Actually, we note
that $H_p^{(L)}\ket{s^{(L)}_{i_1}}
=\left(-\frac{1}{2}\sum_{m=1}^L\epsilon_m+\epsilon_{i_1}\right)\ket{s^{(L)}_{i_1}}
-\sum_{m=1}^{i_1-1}V_{mi_1}\ket{s^{(L)}_m}-\sum_{l=i_1+1}^{L}V_{i_1l}\ket{s^{(L)}_l}$.
It implies the submatrix one of paring Hamiltonian with a simple
construction \beq \label{structurehsub1} H_{\rm
sub1}^{(L)}[i,i]=-\frac{1}{2}\sum_{m=1}^L\epsilon_m+\epsilon_i,\quad
\left. H_{\rm sub1}^{(L)}[i,j]\right|_{i\neq j}=-V_{ij}  \quad
(i,j=1,2,\cdots,L)\eeq Obviously, it is easy to diagonalize the
$H_{\rm sub1}^{(L)}$ for $L = 100\sim 180$ in an ordinary PC
\cite{Our1}. By making use of eq.(\ref{hpaction}) we strictly obtain
the general and explicit form of $H_{\rm sub1}^{(L)}$ and the
arbitrary subspace $H_{{\rm sub}n}^{(L)}$. In this letter, we
consider the typical three kinds of models, that is, the reduced
(constant) model in which $V_{ij}=V$ (constant) for any $i$ and $j$,
the nearest neighbor model in which $V_{ij}=\delta_{|j-i|,1}V$ (only
there is the coupling in the nearest neighbor two pairs) and the
third model which takes $V_{ij}=\beta V/|j-i|$ (the coupling
coefficient decreases pro rate with the ``distance"). They are
denoted respectively by the subscripts RE, NB and RA. Moreover, we
take $V=2\times 10^{-6}$, $\epsilon_m=m V/\lambda$, $\beta=10^{-1}$
and $\lambda$ is a running parameter which can be taken as 10, 20
and $etc$ in the following numerical calculation. Fig.\ref{mypic1}
shows the energy spectrum of subspace 1 of $H_p$ which is obtained
by the direct diagonalization.

Fig.1 and the other figures in this letter are all so-called energy
level figure, in which, every line segment represents an energy
level (or several energy levels whose number is equal to its
degeneration degree), that is, an eigenvalue of Hamiltonian, and
$y$-axis indicates the values of energy levels. In fact, for saving
space, we always put a set of energy level figures together, where
every sub-figure occupies one column and on its top a name of the
considered models (Fig.1) or an ordinal (not including the trivial
ones of $S_0^L$ and $S_L^L$) of subspace (the others) are denoted.
It must be emphasized that in our energy level figures, there are
some energy levels that have not been distinguished clearly since
the limitation of scale of $y$-axis and resolution from printer or
monitor. Actually, it is just the effect we want to display, because
those dense energy levels (large number of energy levels within the
unit scale of energy) form so-called ``energy bands" (the line
segment with some width), two near but obviously separate energy
levels (the difference of two near energy levels has the larger
value) forms so-called ``energy gaps" (two parallel line segment
with an obviously larger interval). For example, it is easy to see
that in the subspace 1 of reduced model has an energy gap standing
at between the energy level $1$ and energy level $2$, and an energy
band including all energy levels except for the energy level $1$.

By the numerical calculating and fitting, analytical continuation
and then theoretical deduce, we can obtain several important and
interesting conclusions.

(1){\em Symmetry in the energy spectrum}: Denoting $E_{\rm
RE,NB,RA}^{(L)}(i,n)$ as the $i$-th energy level of $H_p^{(L)}$ in
the subspace $n$ respectively for the models RE, NB and RA and
arranging as $E^{(L)}(i,n)\leq E^{(L)}(j,n)$ for $i<j$, our data
analysis and numerical fitting indicate that there is the symmetry
between $E^{(L)}(i,n)$ and $E^{(L)}(i,L-n)$. That is \beq
\label{symmetry} E^{(L)}(i,n)=E^{(L)}(i,L-n)- (L +
1)\left(\frac{L}{2} - n\right)\frac{V}{\lambda} \eeq Moreover, this
symmetry is independent of the considered models here and so the
subscript RE or NB or RA are omitted. We think that it can be an
exact theoretical formula since its precision $\leq
10^{-36}$\footnote{We verify the symmetry using a program in
Mathematica ($\copyright$ Copyright 1988-2005 Wolfram Research,
Inc.) and we do not consider the error that might be brought by the
algorithm in Mathematica software.}.
\begin{figure}[h]
\begin{center}
\includegraphics[scale=0.60]{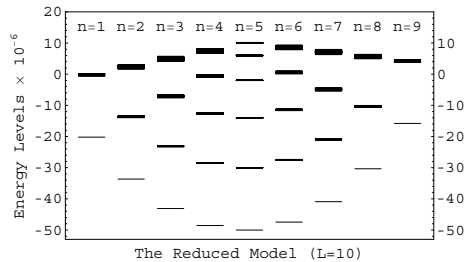}
\vskip -0.1in \caption{The symmetry of energy spectrum, and the
structures of energy gaps and energy bands in the reduced model,
where $\lambda=20$.} \label{mypic2}
\end{center}
\vskip -0.1in
\end{figure}

In fact, a physical symmetry should be a strict result. Newly, we
have finished a strict analytical proof about this property
\cite{OurHS}.

\begin{figure}[h]
\begin{center}
\includegraphics[scale=0.60]{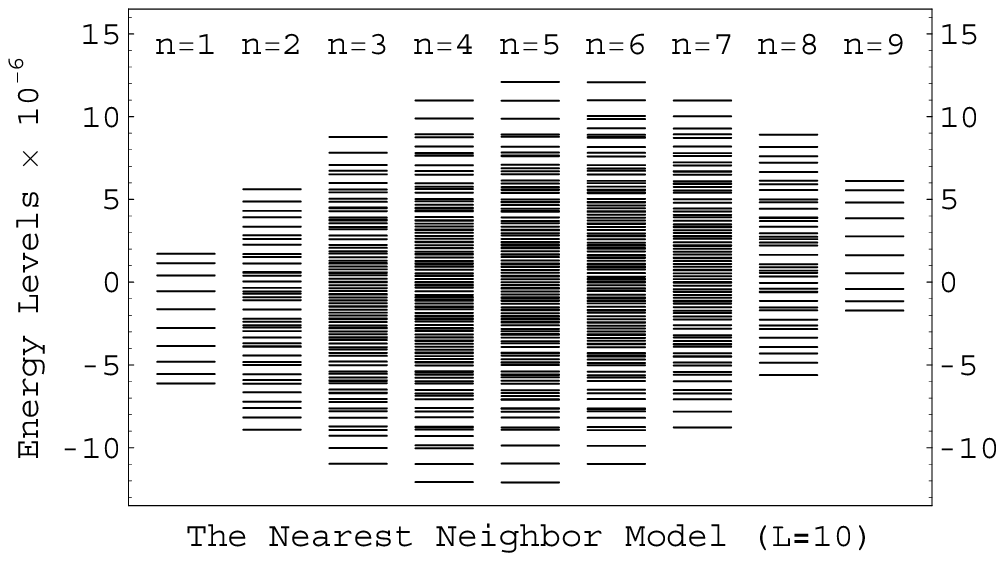} \qquad
\includegraphics[scale=0.60]{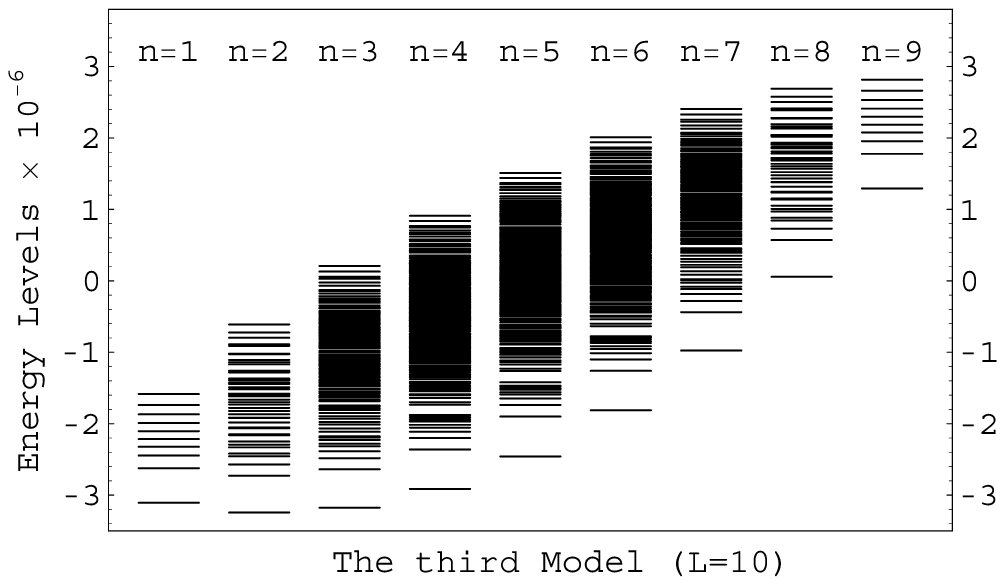}
\end{center} \vskip -0.3in
\caption{The symmetry of energy levels, and the destruction of the
energy gaps and bands in NB and RA models, where
$\lambda=20$.}\label{mypic3} \vskip -0.1in
\end{figure}

(2) {\em Structure of energy gaps and bands}: From the
Fig.\ref{mypic2}, it is easy to see the obvious structure of energy
gaps and bands. Actually, we find that it is a common feature of
energy spectrum in the reduced model. Here, the energy gap(s) means
such a (some) difference(s) between two nearest neighbor energy
levels that it is much larger than the other ones, and the energy
band means such a set of energy levels with very small differences
among the nearest neighbor energy levels, even appearing approximate
degeneration. From the data of energy levels we can conclude that
the number of energy gaps is equal to $n$ (if $n\leq [L/2]$) or
$L-n$ (if $n>[L/2]$) for a given subspace $n$ of $H_p^{(L)}$, and so
is the number of the energy bands since it does not contain a lowest
energy level in this given subspace. The positions of energy gaps
$E^{(L)}_{\rm RE}(k,n)$ appear between $E^{(L)}_{\rm
RE}(L!/((k-1)!(L-k+1)!)+1,n)$ and $E^{(L)}_{\rm
RE}(L!/((k-1)!(L-k+1)!),n)$, where $k=1,2,\cdots,n$ (if $n\leq
[L/2])$ or $k=1,2,\cdots,L-n$ (if $n>[L/2])$). We have verified, in
numerical, the following variation rules for $N\leq 10$: the energy
gaps enlarge with $\lambda$ or $n (\leq [L/2])$ or $L$ increasing,
and lessen with $k$ increasing; the energy bands widen with $k$ or
$L$ increasing, and narrow with $\lambda$ or $n (\leq [L/2])$
increasing. When $n>[L/2]$, the variation of energy gaps and bands
can be known by the symmetry in energy spectrum. However,
Fig.\ref{mypic3} shows that the nearest neighbor model destroys the
structure of energy gaps and energy bands, and so almost does the
third model except for the first energy gap. The destruction
strength of the third model for the first energy gap depends on
variety strength of the coupling coefficients with the distance
``$|j-i|$", the more rapidly the coupling coefficients decrease with
the distance, the more largely the first energy gap is destroyed.
Fig.\ref{mypic3} gives out an example displaying these features.

(3) {\em Lowest energy levels}: In the reduced model, we find that
the lowest energy level (LEL) of $H_p^{(L)}$ is at the half filling
subspace $[L/2]$. In terms of the theoretical analysis on
Hamiltonian and numerical fitting for the energy levels, we
conjecture that the LEL of $H_p^{(L)}$ can be estimated by the
following  formula with a large leading term \beqa \label{lelhf}
E^{(L)}_{\rm
RE}\left(1,\left[\frac{L}{2}\right]\right)&\approx&-\left[\frac{L}{2}\right]
\left[\frac{L+1}{2}\right] V - 2
\log\left(\left[\frac{L-1}{2}\right]\right)\nonumber\\
& &\left(\sin^2\left(\frac{L\pi}{2}\right)
\frac{V}{\lambda}+\cos^2\left(\frac{L\pi}{2}\right)\frac{V}{\lambda^2}\right)
\eeqa From the fact that $E_{\rm RE}^{(L)}(1,n)\approx E_{\rm
RE}^{(L)}(1,n-1)+(E_{\rm RE}^{(L)}(1,[L/2])-E_{\rm
RE}^{(L)}(1,1))/([L/2]-1)-([L/2]-2(n-1))V$ obtained by the data
analysis and numerical fitting, it follows that \beqa E_{\rm
RE}^{(L)}(1,n)&\approx& \left(1-\frac{n-1}{[L/2]-1}\right)E_{\rm
RE}^{(L)}(1,1)-\frac{(n-1)}{[L/2]-1}\nonumber\\
& &\left(\left[\frac{L}{2}\right]
\left[\frac{L+1}{2}\right]+([L/2]-1)([L/2]-n)\right)V\nonumber\\[8pt]
& &-\frac{2(n-1)\log([(L-1)/2])}{[L/2]-1}
\left(\sin^2\left(\frac{L\pi}{2}\right)
\frac{V}{\lambda}+\cos^2\left(\frac{L\pi}{2}\right)\frac{V}{\lambda^2}\right)\nonumber\\
&=& \left(1-\frac{n-1}{[L/2]-1}\right)E_{\rm
RE}^{(L)}(1,1)-\frac{(n-1)}{[L/2]-1}\nonumber\\
& &\left(\left[\frac{L}{2}\right]\left(
\left[\frac{L+1}{2}\right]+\left[\frac{L}{2}\right]\right)-
n\left(\left[\frac{L}{2}\right]-1\right)-\left[\frac{L}{2}\right]\right)V\nonumber\\[8pt]
& &-\frac{2(n-1)\log([(L-1)/2])}{[L/2]-1}
\left(\sin^2\left(\frac{L\pi}{2}\right)
\frac{V}{\lambda}+\cos^2\left(\frac{L\pi}{2}\right)\frac{V}{\lambda^2}\right)\nonumber\eeqa
It implies that \beqa\label{leln} E_{\rm RE}^{(L)}(1,n)&\approx &
\left(1-\frac{n-1}{[L/2]-1}\right)E_{\rm
RE}^{(L)}(1,1)+(n-1)\left(n-(L-1)\frac{[L/2]}{[L/2]-1}\right)V\nonumber\\[7pt]
& &-\frac{2(n-1)\log([(L-1)/2])}{[L/2]-1}
\left(\sin^2\left(\frac{L\pi}{2}\right)
\frac{V}{\lambda}+\cos^2\left(\frac{L\pi}{2}\right)\frac{V}{\lambda^2}\right)\eeqa
which can be used to estimate the LEL $E_{\rm RE}^{(L)}(1,n)$ in a
given subspace $n\leq [L/2]$. For the other subspaces ($n>[L/2]$),
we can estimate their LEL by eq.(\ref{symmetry}). When $L=10,
\lambda=20$, the absolute errors of the estimated values for all of
subspaces from 1 to $[L/2]$ are $\leq 10^{-9}$ and their relative
errors are $\leq 10^{-5}$.

\begin{figure}[h]
\begin{center}
\includegraphics[scale=0.60]{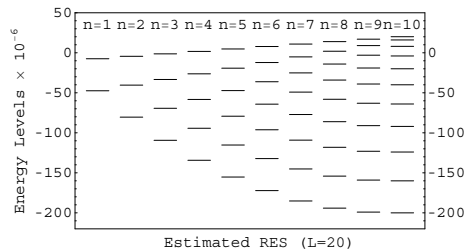}
\end{center}
\vskip -0.3in \caption{The estimated the lowest energy level and
representative energy spectrum ($\lambda=20$) for the subspaces from
$n=1$ to $n=10$} \label{mypic4} \vskip -0.1in
\end{figure}

(4) {\em The representative energy levels and energy spectrum}:
Since the typical structure of energy gaps and energy bands as well
as the large density of energy levels for a given band in the
reduced model, we can introduce a so-called representative energy
level (REL) which is defined by an average of all of energy levels
in a given energy band, that is \beqa\label{rel} E^{(L)}_{\rm
r}(k,n)&=&\frac{1}{L!/\left(k!(L-k)!\right)-L!/\left((k-1)!(L-k+1)!\right)}\nonumber\\
& &
\sum_{i=L!/\left((k-1)!(L-k+1)!\right)+1}^{L!/\left(k!(L-k)!\right)}E^{(L)}_{\rm
RE}(i,n) \eeqa where $k=1,2,\cdots,n$ (if $n\leq [L/2])$ or
$k=1,2,\cdots,L-n$ (if $n>[L/2])$). This REL can be used to indicate
the corresponding energy band, in particular, in the interesting
half filling case since it with narrower energy bands. Obviously, in
a given subspace $n$, the structure features of energy spectrum of
pairing model can be represented by the simple representative energy
spectrum which consists of a LEL and $n (\leq [L/2])$ or $L-n
(n>[L/2])$ RELs. In particular, based on the data analysis and
numerical fitting for the RELs, we find that the widths of
representative energy gaps, that is, $\Delta_{\rm
r}^{(L)}(k,n)=E_{\rm r}^{(L)}(k,n)-E_{\rm r}^{(L)}(k-1,n)$ (where
$E_r^{(L)}(0,1)=E_{\rm RE}^{(L)}(1,1)$) from low to high decrease
near a constant $2V$, which means that $\Delta_{\rm
r}^{(N)}(k,n)\approx\Delta_{\rm r}^{(N)}(k-1,n)-2V$, and
$\Delta_{\rm r}^{(L)}(k,n)\approx \Delta_{\rm r}^{(L)}(k,n+1)$. From
them we have \beqa E_r^{(L)}(k,n)&\approx&\Delta_{\rm
r}^{(L)}(k,n)+E_r^{(L)}(k-1,n)+(\mbox{high order approximation})\nonumber\\
&=&\sum_{i=1}^k \Delta_{\rm
r}^{(L)}(i,n)+E_r^{(L)}(0,n)+(\mbox{high order approximation})\nonumber\\
&=&\sum_{i=1}^k \Delta_{\rm
r}^{(L)}(i,n)+E_{\rm RE}^{(L)}(1,n)+(\mbox{high order approximation})\nonumber\\
&=& k \Delta_{\rm
r}^{(L)}(1,n)-k(k-1)V+E_r^{(L)}(1,n)+(\mbox{high order approximation})\nonumber\\
&=& k \Delta_{\rm r}^{(L)}(1,1)-k(k-1)V+E_r^{(L)}(1,n)+ (\mbox{high
order approximation}) \nonumber\eeqa Adding two additional
compensatory terms (high approximation) to eliminate the error since
the many times recursions $$ (\mbox{high order
approximation})\approx \frac{2\;
\delta_{k,[L/2]}}{[L/2]-1}\frac{V}{\lambda}-
\sin^2\left(\frac{(n-k)\pi}{2}\right)\frac{V}{\lambda^2}$$ Of course
, the form of this two additional terms is given by the numerical
fitting. Therefore, we obtain the formula to estimate the RELs in
the subspace $n$ ($n\leq [L/2]$) by \beqa \label{reln}
E_r^{(L)}(k,n)& \approx & k \Delta_{\rm
r}^{(L)}(1,1)+\left(1-\frac{n-1}{[L/2]-1}\right)E_{\rm
RE}^{(L)}(1,1)\nonumber\\[7pt]
& &+\left((n-1)n-(n-1)(L-1)\frac{[L/2]}{[L/2]-1}-k(k-1)\right)V
\nonumber\\[7pt]
& & - \frac{2(n-1)\log([(L-1)/2])}{[L/2]-1}\left(
\sin^2\left(\frac{L\pi}{2}\right)\frac{V}{\lambda} +
\cos^2\left(\frac{L\pi}{2}\right)\frac{V}{\lambda^2}\right)\nonumber\\[7pt]
& & + \frac{2\; \delta_{k,[L/2]}}{[L/2]-1}\frac{V}{\lambda}-
\sin^2\left(\frac{(n-k)\pi}{2}\right)\frac{V}{\lambda^2}\eeqa where
$\Delta_{\rm r}^{(L)}(1,1)=E^{(L)}_r(1,1)-E_{\rm RE}^{(L)}(1,1)$,
which can be calculated by the directly diagonlizing the subspace 1
of $H_p$. Note that we, based on the data analysis and numerical
fitting, add two little addition terms in order to include the
contribution from high order approximations among the representative
energy gaps . When $L=10, \lambda=20$, the absolute error of this
estimation is $\sim 10^{-9}$ and its relative error is $\sim
10^{-3}$. Then, based on the symmetry in the energy spectrum, we can
obtain the values of representative energy spectrum in the other
subspaces. Fig.\ref{mypic4} is an example which is estimated by our
above scheme.
\noindent The representative energy spectrum in subspaces $11$ to
$19$ has not been pictured since their shapes can be known by the
symmetry of energy spectrum. For the subspace 2, we can calculate
out the absolute errors of one estimated the lowest energy level and
two estimated representative energy levels are respectively
$5.80231\times 10^{-9}$, $2.63859\times 10^{-9}$ and $6.47206\times
10^{-9}$, their relative errors are respectively $0.0000721591$,
$0.0000653016$ and $0.00147345$. Therefore, the lowest energy levels
and the representative energy levels can have the better and usable
precision in numerical calculation as well as estimation by our
method.

It is clear that we have solved the lowest energy levels of all of
the subspaces and representative energy spectrum of the pairing
model in numerical by our direct diagonalization of Fock space.
Moreover, we show the structure of the energy levels and energy gaps
in the pairing model for the reduced model. After finishing the
proof of symmetry between the energy spectrum of pairing model, we
would like to show the origin of leading term of the LEL of $H_p$ in
the near. In principle, our method should be able to extend to some
similar spin Hamiltonian systems. Of course, because our estimated
formula are obtained by the numerical fitting and analytical
continuation, it is still a problem how large $N$ our scheme is
suitable to. In addition, for the width of energy bands we have not
found a good estimated method yet. More knowledge about the other
models needs to explore.

We specially acknowledge interesting discussions with Feng Xu and
valuable suggestions by her. We are grateful to Ningbo Zhao,
Xiaodong Yang, Xiaosan Ma, Hao You, Wanqing Niu, Rengui Zhu and
Xiaoqiang Su for our cooperations in the Quantum Theory Group, the
Institute for Theoretical Physics of University of Science and
Technology of China. This work was funded by the National
Fundamental Research Program of China under No. 2001CB309310, and
partially supported by the National Natural Science Foundation of
China under Grant No. 60573008.



\begin{references}
\bibitem{bcs}
J. Bardeen, L. N. Cooper, and J. R. Schrieffer Phys. Rev. {\bf
108}{(1957)} {1175}
\bibitem{report} von Delft, J. and D.C. Ralph, Phys. Rep.
{\bf 345} (2001) 61 and its refs.
\bibitem{dukelsky}J. Dukelsky and G. Sierra, Mod. Phys. Rev. {\bf
76} (2004) 643 and its refs.
\bibitem{richardson}R. W. Richardson., Phys. Lett. {\bf 3} (1963)
277; Phys. Lett. {\bf 5} (1963)82
\bibitem{Dukelsky2} J. Dukelsky and G. Sierra
Phys. Rev. Lett. {\bf 83} (1999)172
\bibitem{Amico1}L. Amico and A.
Osterloh Phys. Rev. Lett. {\bf 88} (2002)127003
\bibitem{Amico2}L.
Amico, A. Di Lorenzo, and A. Osterloh Phys. Rev. Lett. {\bf 86}
(2001)5759
\bibitem{Volya}A. Volya, B. A. Brown, and V. Zelevinsky, Phys.Lett. B {\bf 509} (2001)37
\bibitem{simulation} L.-A. Wu, M. S. Byrd, and D. A. Lidar, Phys.
Rev. Lett. {\bf 89} (2002) {057904}, Phys. Rev. Lett., {\bf 90}
{(2003)} {249804}; J. Dukelsky, J. M. Rom\'{a}n, and G. Sierra, Phys
Rev Lett., {\bf 90} {(2003)} {249803}
\bibitem{Burglin}O.Burglin and N.Rowley, Nucl. Phys. A {\bf
602}(1996)21
\bibitem{Molique}H. Molique and J. Dudek, Phys. Rev. C {\bf
56}(1997)1795
\bibitem{Ourqst}An Min Wang and Xiaodong Yang, Phys. Lett. A, {\bf 352}(2006) 304
\bibitem{Ourqse}Xiaodong Yang, An Min Wang, Feng Xu, and Jiangfeng
Du, Chemical Physics Letters, {\bf 422} (2006) 20
\bibitem {Our1}Xu Feng, An Min Wang, Xiaodong Yang, Xiaosan Ma, and Hao
You, Commun. Theor. Phys. {\bf 44} (2005) 171
\bibitem{OurHS}An Min Wang and Ren Gui Zhu, Chinese Phys. Lett. {\bf
23} (2006) 2542
\end{references}
\end{document}